\documentclass[10pt]{article}
\usepackage{amsmath,amssymb}
\usepackage[dvips]{epsfig}
 \allowdisplaybreaks
 \numberwithin{equation}{section}

\def\be{\begin{equation}}
\def\ee{\end{equation}}
\newcommand{\rf}[1]{(\ref{#1})}

\def\a{{\sf a}}
\def\N2{${\cal N}=2$}
\def\4N{${\cal N}=4$}
\def\F{{\mathcal F}}
\def\d{\partial}
\def\half{{\textstyle{\frac12}}}
\def\ha{\frac12}

\def\res{{\rm res}}

\def\Im{\mathop{\rm Im}\nolimits}

\begin{document}
\begin{flushright}
FIAN/TD-05/06\\
ITEP/TH-15/06\\
MPIM-44/06
\end{flushright}
\vspace{1.0 cm}

\setcounter{footnote}{3}
\renewcommand{\thefootnote}{\fnsymbol{footnote}}
\begin{center}
{\Large\bf WDVV equations for 6d Seiberg-Witten theory\\
and bi-elliptic
curves}\\
\vspace{1.0 cm} {\Large H.W.Braden\footnote{School of Mathematics,
University of Edinburgh, Edinburgh, Scotland; e-mail: hwb@ed.ac.uk }, A.Marshakov\footnote{Lebedev Physics Institute and ITEP, Moscow, Russia; e-mail: mars@lpi.ru, mars@itep.ru},
A.Mironov\footnote{Lebedev Physics Institute and ITEP, Moscow Russia;
e-mail: mironov@lpi.ru, mironov@itep.ru}, A.Morozov\footnote{ITEP,
Moscow, Russia; e-mail: morozov@itep.ru}
}\\
\vspace{0.6 cm}
\end{center}
\begin{quotation} \noindent We present a generic derivation of
the WDVV equations for 6d Seiberg-Witten theory, and extend it to
the families of bi-elliptic spectral curves. We find that the
elliptization of the naive perturbative and nonperturbative 6d
systems roughly ``doubles" the number of moduli describing the
system.
\end{quotation}

\renewcommand{\thefootnote}{\arabic{footnote}}
\setcounter{section}{0} \setcounter{footnote}{0}
\setcounter{equation}0

\section{Introduction: WDVV equations and residue formulas}

Complex geometry is playing an increasingly important role in
non-perturbative physics. For example, modern (topological) string
theory, including Seiberg-Witten (SW) theory, intensively exploits
the prepotentials of complex manifolds, the (generalized) period
matrices of which appear as couplings in the associated
low-energy, effective field theory Lagrangians. In many cases
these prepotentials satisfy some particularly nice non-linear
differential equations, and these may be integrated using the
basic properties of the underlying complex geometry. The general
theory of these equations is still far from being complete. This
paper will focus on one such class of equations, the WDVV
equations, and provide some new solutions to these arising from
six dimensional (6d) Seiberg-Witten theory. We begin by reviewing these
equations.

In many known cases the nontrivial part of the complex geometry
effectively reduces to families of one-dimensional complex
manifolds, or curves, and both the equations and the curves are
related to parts of well-known infinite-dimensional integrable
hierarchies. The latter may be further effectively rewritten in
group-theoretical terms, or even ``linearized" in the form of the
Virasoro/W-algebra constraints. Other ingredients of this picture
suggest various universal properties, presumably generalizable
beyond one complex dimension. This is particularly true of the
relations satisfied by the generalized period matrices, including
their first derivatives; equivalently the third derivatives of
prepotentials. Such third derivatives have the sense of
three-point functions in string theory and are very robust objects
because of the large (three-dimensional) group of automorphisms of
the world-sheet spheres. The basic relations satisfied by the
third derivatives are known as the WDVV equations \cite{WDVV} and
these were originally obtained from the crossing-symmetry of the
four-point functions in topological string theory.

In their most general form \cite{MMM} the WDVV equations can be
presented as a system of algebraic relations
\begin{equation} \label{WDVV}
\F_I{\F}_J^{-1}\F_K = \F_K{\F}_J^{-1}\F_I, \ \ \ \ \ \ \forall\
I,J,K \end{equation} for the matrices of third derivatives
\begin{equation} \label{matrF}
\|{\F}_{I}\|_{JK}= \frac{\d^3\F}{\d T_I\,\d T_J\,\d T_K}
\equiv\F_{IJK} \end{equation}
of some function $\F (T_K)$. %
Originally one considered only a particular class of solutions to
the system (\ref{WDVV}), where one of the matrices $\F_{I_0}$ was
a constant (matrix) independent of $\{ T_J\}$, see \cite{WDVV,Dub}.
This restriction corresponded to the existence of a distinguished
vector (the vacuum) in the space of states of topological string
theory with corresponding parameter $T_{I_0}$, the ``cosmological
constant". In the framework of SW theory however, there is no
natural place for such a constraint, see \cite{MMM,forms,more,MM};
moreover, this constraint
violates the basic symmetries of the SW theory, like electric-magnetic
duality \cite{dWMa}.
In fact, it turns out that this constancy
condition is inessential to the proof of solutions to the WDVV
equations given by prepotentials of complex manifolds for which
the third derivatives are expressed by a {\em residue formula}.

The residue formula relevant for the WDVV equations takes the
following form \cite{KriW}. One has a Riemann surface $\Sigma$
endowed with a meromorphic generating one-form $dS= -zd\tilde z$.
Then
\begin{equation} \label{residue} \frac{\d^3 \F}{ \d T_I\d T_J\d T_K} =
\res_{d\tilde z=0}\left(\frac{d\Omega_Id\Omega_Jd\Omega_K}{ dz
d\tilde z}\right) =
\sum_{\alpha}\res_{z_\alpha} \left(\frac{\phi_I\phi_J\phi_K}{
{d\tilde z /dz}}dz\right) =
\sum_{\alpha}\frac{\phi_I(z_\alpha)\phi_J(z_\alpha)
\phi_K(z_\alpha)}{ \Psi_\alpha}, \end{equation} where $d\tilde z =
\Psi_\alpha\cdot(z-z_\alpha)dz + \ldots$ as $z\to z_\alpha$. The
set of one-forms $\{ d\Omega_I\}$ corresponds to the set of
parameters $\{ T_I\} $ via \begin{equation} \label{sdiff} \frac{\d
dS}{ \d T_I} = d\Omega_I \equiv \phi_I dz. \end{equation} With
such a residue formula the proof of the WDVV equations
(\ref{WDVV}) reduces to solving a system of linear equations
\cite{BMRWZ,MaWDVV}, the solution of which requires only two
conditions to be fulfilled:

\begin{itemize}

\item[1.] A matching condition between the number of deformation
parameters (moduli) $n_m=\#(I)$ and the number of critical points
$n_z=\#(\alpha )$ in the residue formula
\begin{equation}
\label{matching}
n_z=n_m
\end{equation}

\item[2.] Nondegeneracy of the matrix $\phi_{I}(z_\alpha )$,
\begin{equation}
\label{detW}
\det_{I\alpha }\| \phi_{I}(z_\alpha
)\| \neq 0
\end{equation}
This is believed to always be fulfilled in ``general position".
\end{itemize}

Supposing these conditions to be satisfied the structure constants
$C_{IJ}^K$ of the associative algebra underlying the WDVV equations
may be found from the system of linear equations (one for each
$z_\alpha$)
\begin{equation} \label{eqc} \phi_I(z_\alpha )\phi_J(z_\alpha )
=\sum_K C^K_{IJ}(\xi)\phi_K(z_\alpha )\cdot\xi(z_\alpha ).
\end{equation}
This algebra is isomorphic to the more usual one considered in
relation to the WDVV equations provided $\xi(z_\alpha )\ne0$,
which we henceforth assume\footnote
{An alternative approach to this is to
construct an associative algebra of forms, see
\cite{forms,more,mir}.}. Utilising the matching and nondegeneracy
assumptions we may solve (\ref{eqc}) to give
\begin{equation} \label{litc} C^K_{IJ}(\xi) = \sum_\alpha
\frac{\phi_I(z_\alpha )\phi_J(z_\alpha )}{\xi(z_\alpha )}
\left(\phi_K(z_\alpha )\right)^{-1}.
\end{equation}
The WDVV equations follow from the associativity of (\ref{eqc})
once we establish the consistency of the relation
\begin{equation} \label{feta} {\mathcal
F}_{IJK} = \sum_L C_{IJ}^L(\xi)\eta_{KL}(\xi), \end{equation}
which expresses the structure constants in terms of the third
derivatives. Here we have introduced a ``metric''
\begin{equation}
\label{metric} \eta_{KL}(\xi) = \sum_M \xi_M {\mathcal F}_{KLM}
,
\end{equation}
where the parameters $\xi_M$ are arbitrary, subject to $\eta_{KL}$
being invertible. Upon defining the differential $\xi(z)dz$ with
$$\xi(z)=\sum_M \xi_M \phi_M(z),$$
one sees that the values of $\xi(z_\alpha)$ in (\ref{eqc})
determine $\xi_M$ and visa versa using $$\xi_M =\sum_\alpha
\phi_M(z_\alpha)\sp{-1}\xi (z_\alpha).$$ The consistency of
(\ref{feta}) now follows simply if ${\mathcal F}_{KLM}$ are given
by a residue formula (\ref{residue}):
\begin{align*}
\sum_K C_{IJ}^K(\xi)\eta_{KL}(\xi) &= \sum_{K,\alpha ,\beta }
\frac{\phi_I(z_\alpha )\phi_J(z_\alpha )}{\xi(z_\alpha )}\cdot
\left(\phi_K(z_\alpha )\right)^{-1}\cdot\phi_K(z_\beta )
\phi_L(z_\beta )\xi(z_\beta )\Psi^{-1}_\beta
\\
&= \sum_\alpha \frac{\phi_I(z_\alpha )\phi_J(z_\alpha
)}{\xi(z_\alpha )} \phi_L(z_\alpha )\xi(z_\alpha )\Psi^{-1}_\alpha
=
\sum_\alpha  \Psi^{-1}_\alpha \phi_I(z_\alpha )\phi_J(z_\alpha
)\phi_L(z_\alpha )\\
 &= {\mathcal F}_{IJL}.
 \end{align*}

\section{Seiberg-Witten prepotentials}

We shall now construct a new class of SW prepotentials \cite{SW} and
solutions to the WDVV equations using the general formalism
described in the introduction. The focus of our attention will be
on curves ${\Sigma
}\subset\mathcal{E}_\tau\times\mathcal{E}_{\tilde\tau}$ lying in
the product of two elliptic curves,
\begin{equation} \Sigma: \ \ \ {\mathcal H}(\tilde z,z) = 0, \ \ \ dS=
-zd\tilde z \label{sepv}.
\end{equation} Here ${\cal H}$ is doubly-periodic in both $z$ and $\tilde z$
with respective periods $(1,\tau)$ and $(1,{\tilde \tau})$. Such
curves (\ref{sepv}) are endowed with a ``symmetric" generating
differential: under exchange of the two tori this becomes a Legendre
transform. In particular cases the $\tilde z$-torus will be taken to
degenerate with $\tilde\tau\to +i\infty$. Then it is convenient to
use the co-ordinate on a cylinder $\tilde z \rightarrow \log w$
instead of $\tilde z$ itself. In this case the generating
differential becomes
\begin{equation}
\label{dS6d} dS = -z\frac{dw}{ w}
\end{equation}
and acquires the form more commonly appearing in SW theory (see
\cite{SWbooks,GMrev}, and references therein). Our setting corresponds to a
6d gauge theory with two extra dimensions compactified
onto the $z$-torus.

The variation of the generating differential (\ref{sepv}) may be
written as \begin{equation} \label{vardSg} \delta(dS) =
\delta{\cal H}\frac{dz}{{\cal H}'_{\tilde z}}\ \in\
\bigoplus_{I=1}^{n_m} d\Omega_I ,\end{equation} where the
right-hand side is understood as a linear combination of all {\em
canonical} differentials. The most common basis of the space of
all canonical differentials consists of the holomorphic or Abelian
differentials of the first kind $\{ d\omega_i\}$, and the
meromorphic or Abelian differentials of the second and third
kinds. If $\{A_i,B_i\}\subset H_1(\Sigma)$ are a canonical homology
basis for our
curve $\Sigma$ the holomorphic differentials may be used to vary
the $A_i$-periods of $dS$ while the meromorphic differentials may
be used to describe any poles of $dS$ and any monodromy, or jumps,
it may have. (Note that $z$ and $\tilde z$ are non-single valued
Abelian integrals and not functions on $\Sigma$.) Abelian
differentials of the third kind, $d\Omega_{P_+,P_-}$, with
residues $\pm1$ at $P_\pm$, also arise when allowing degenerations
and a handle is shrunk to a pair of marked points ($P_\pm$). These
various sorts of differentials will correspond in (\ref{vardSg})
to different variations of the parameters of ${\mathcal H}$.

In this paper we will focus on the case when the variations
(\ref{vardSg}) are {\em almost all} accounted for by the
holomorphic differentials (or, when there is degeneration, Abelian
differentials of the third kind).
However, it turns out, that if we want to restrict ourselves to
only this class of generalized holomorphic differentials, there is
always a mismatch by one with the matching condition
(\ref{matching}) needed for the validity of the WDVV equations.
For us this mismatch is filled by a variation in $\tau$ of $dS$,
producing in (\ref{vardSg}) a particular differential
$d\Omega_\tau$ with jump along the $B_i$-cycles that nontrivially
project to the $B$-cycle of the base $z$-torus. Thus we are
considering variations $\delta(dS)$ of the form (for some point
$P_*$)
\begin{equation}
\delta(dS)=\sum_{i=1}\sp{g_\Sigma}\,\delta \a_i
d\omega_i+\sum_{j}\delta a_j\, d\Omega_{P_*,P_j}+\delta\tau
d\Omega_\tau,
\end{equation} and corresponding ``times''
$\{T_K=\a_i,a_k,\tau \}$
\begin{equation} \label{per} \{T_K\}:\ \ \a_i =
\oint_{A_i} dS,\ \ \ a_k = -\res_{P_k} dS,\ \ \ \tau = \oint_B dz.
\end{equation}Then we define (up to a
constant) the function $\mathcal{F}(T_K)$ in terms of its
derivatives $\{ \frac{\d\F}{\d T_K} \}$ by
 \begin{equation}
\label{grad}    \frac{\d\F}{ \d\a_i} = \oint_{B_i} {\tilde z} dz,\ \
\ \frac{\d\F}{ \d a_k} = \int_{P_k}^{P_*} {\tilde z} dz,\ \ \
\frac{\d\F}{\d\tau} =
\oint_A {\tilde z}dS.
\end{equation}
The integrability of (\ref{grad}) and so the existence of a
prepotential $\mathcal{F}(T_K)$ now follows from the Riemann
bilinear identities satisfied by the differentials on a Riemann
surface,
\begin{equation} \label{rbi} \int_\Sigma
d\Omega_I\wedge d\Omega_J = 0.
\end{equation}
Considering the canonical holomorphic differentials $\{
d\omega_i\}\subset \{ d\Omega_I\}$, for example, these identities
ensure that the period matrix $T_{ij}=\partial\sp2\F/\d\a_i\d\a_j$
of $\Sigma$ is symmetric,
\begin{equation} \label{Rid}
T_{ij}=\oint_{B_i} d\omega_j = \oint_{B_j} d\omega_i=T_{ji},
\qquad i,j=1\ldots g_\Sigma.
\end{equation} Likewise consideration of the whole set of differentials
satisfying (\ref{rbi}) similarly shows that the generalized period
matrices (\ref{Rid}) can be integrated to yield a (locally defined)
function $\F (T_K)$ \cite{KriW} of the periods, residues and jumps
of the generating differentials. In addition to (\ref{Rid}) one
gets, for example, that
\begin{equation}
\label{rbiaa} \frac{\d^2\F}{\d a_k\d\a_i} =
\oint_{B_i}d\Omega_{P_k,P_*} = \int_{P_k}^{P_*} d\omega_i =
\frac{\d^2\F}{\d\a_i\d a_k}
\end{equation}
and
\begin{equation}
\label{rbit}
\frac{\d^2\F}{\d\tau\d\a_i} = 
\oint_{B_i}d\Omega_\tau = 
\oint_{A} \tilde z\, d\omega_i = \frac{\d^2\F}{\d\a_i\d\tau}.
\end{equation}At this juncture we simply note (to be elaborated upon
below) that ``contact terms", or the values of Abelian integrals
at the intersection points $P_k=A_k\cap B_k$, appear in our
setting as a consequence of the multi-valuedness of the generating
differential $dS$.

In what follows we are going to apply these general formulae to
particular examples of SW prepotentials. We start with the so
called 6d supersymmetric QCD and  then generalize it
to the case of the curves (\ref{sepv}).

\subsection{The perturbative 6d case}

We will now consider the perturbative prepotential of 6d supersymmetric QCD. By 6d supersymmetric
QCD we follow \cite{XYZ,MaMi} and mean the \N2 SUSY
four-dimensional gauge theory with $SU(N)$ gauge group and
$N_f=2N$ fundamental matter multiplets together with the extra
Kaluza-Klein modes corresponding to adding {\em two} compact
dimensions. All $N_f=2N$ matter multiplets are taken throughout
with vanishing masses. The perturbative prepotential of this
hypothetical 6d theory compactified on the torus
$\mathcal{E}_\tau=\mathbb{C}/(1,\tau)$ was calculated in
\cite{MaMi} using the residue formula (\ref{residue}). Let
\begin{equation} \label{tauto} x = \wp(z|\tau), \ \ \ y =
-\half\wp'(z|\tau), \qquad y^2 = \prod_{i=1}^3(x-e_i),
\end{equation}
be standard (affine) coordinates for the torus $\mathcal{E}_\tau$.
In terms of these the curve $\Sigma$ is defined by
\begin{equation}
\label{6dc} w = e^u \prod_{j=1}^{N}\frac{\theta_1(z-a_j)}{
\theta_1(z)} = P(x) + y\,Q(x), \ \ \ \ \sum_{j=1}^{N} a_j = 0.
\end{equation}
It is endowed with the generating differential (\ref{dS6d}). The
perturbative curve (\ref{6dc}) defines an elliptic function
$w\in\mathbb{C}^*$ on the torus and so $g_\Sigma=1$. The
prepotential is computed as a function of the degenerate SW periods
or residues
\begin{equation} \label{resSW} a_j = 
-\res_{P_j}
dS = 
\res_{z=a_j}  z\frac{dw}{ w}, \ \ \ \ j=1,\dots,N .\end{equation}
Choosing a set of $N-1$ independent quantities from the $N$
variables $a_j$ subject to $\sum_{j=1}^{N} a_j = 0$ in a standard
way, say $a_j\to a_j-a_{N}$, the resulting prepotential is easily
written in terms of the quantum tri-logarithm function \cite{MaMi}.

In addition to the variables (\ref{resSW}) there are two more
natural parameters in (\ref{6dc}). There is the modulus $\tau$ of
elliptic curve, the complexified ratio of the two compactification
radii of the 6d theory, together with the coefficient of
proportionality $\exp(u)$ which is ``reminiscent" of the scale
factor of the 6d theory and related to the coupling constant of
the ``microscopic" gauge theory. (Here we are writing the
coefficient of proportionality of \cite{MaMi} in the exponential
form of \cite{str}.)

The residue formula (\ref{residue}), used in \cite{MaMi}, can be
extended to include these extra parameters. Let
$$\a =
\oint_{A} dS, \qquad a_k = -\res_{P_k} dS,\qquad \tau = \oint_B
dz,$$ (recalling that we are in the genus one setting, so $A_1=A$,
$\a_1=\a$ here). To apply the general formalism we need to relate
$u$ to the period $\a$. The delicate point here is that $dS$ is
\emph{not} single-valued. Unlike the usual SW setting we now need
to specify a point $P_0=A\cap B$ and fix some $z_0=z(P_0)$ and
$w_0=w(P_0)$ \cite{KriW}; the prepotential depends on the choice
of homology cycles and various ``contact terms'' must be included.
As
$$dS=-z\frac{dw}{w}=dz\log w - d\left(z\log w\right)$$
then
\begin{align}\a&=-\int_{z_0}\sp{z_0+1}d\left(z\log w\right)+\oint_{A}\log w dz
=-\log w_0+\oint_{A}udz+\sum_{j=1}\sp{N}\oint_{A}
\log\frac{\theta_1(z-a_j)}{ \theta_1(z)}dz \label{auwt}\\
&=-\log w_0+u.\label{auw}\end{align} To see that the final term
of (\ref{auwt}) may be taken to vanish we use the identity
$$
\log\frac{\theta_1 \left(z-a_j\right)}{ \theta_1
\left(z\right)}= \log\frac{\sin \pi(z-a_j)}{ \sin \pi
z}-
4\sum_{n=1}\sp{\infty}\frac{1}{n}\frac{q\sp{2n}}{1-q\sp{2n}}\sin\pi
n(2z-a_j)\sin \pi n a_j.$$ Then by periodicity the trigonometric
terms vanish upon integration and
$$\sum_{j=1}\sp{N}\oint_{A}
\log\frac{\theta_1(z-a_j)}{ \theta_1(z)}dz=
\sum_{j=1}\sp{N}\oint_{A} \log\frac{\sin \pi(z-a_j)}{ \sin \pi
z}dz.$$ The value of this last integral depends on the choice
of contour $A$: if it is such that $\Im a_j < \Im z_0$ ($j=1,\dots,
N$) then the contour may be slid to infinity and the integrals
vanish, while other choices of contour will differ by an integer
multiple of $2\pi i$. Thus for our chosen homology basis we obtain
(\ref{auw}) showing that (for constant $w$) variations in $\a$ and
$u$ are the same. Following from
\begin{equation}
\label{variat} \delta\log w = \delta u + \delta z\,\frac{d\log w}{
dz} - \sum_i\delta a_i \frac{\theta_1'(z-a_i)}{\theta_1(z-a_i)} +
\delta\tau
\left(\sum_{j=1}^{N}\frac{\theta_1''(z-a_j)}{\theta_1(z-a_j)}
-N\frac{\theta_1''(z)}{\theta_1(z)}\right) \end{equation} to each
of the times $\{ T_I = u,a_j,\tau\} $ (at constant $w$) we may
associate the differentials $\{ d\Omega_I =
dz,d\Omega_j,d\Omega_\tau\}$ via
\begin{align}\label{u} \frac{\d dS}{\d u} &= dz,\\
 \label{sj}\frac {\d dS}{ \d a_j} &= d\Omega_j=
\left(\frac{\theta_1'(z-a_{N})}{\theta_1(z-a_{N})}
-\frac{\theta_1'(z-a_j)}{\theta_1(z-a_j)}\right)dz, \ \
\ j=1,\dots,N-1,\\
\label{stau} \frac{\d dS}{ \d \tau} &= d\Omega_\tau.
\end{align} Thus (\ref{u}) gives us the (unique) holomorphic
differential on the torus. The Abelian differentials of the
third-kind (\ref{sj}) used in \cite{MaMi} can also be expanded
over the basis of
\begin{equation} \label{hydeg} \frac{dz}{
w}\left(1,\dots,\wp(z)^{[N/2]}\right) = \frac{dx}{
wy}\left(1,\dots,x^{[N/2]}\right),\  \text{ and}\ \ \frac{dz}{
w}\left(1,\dots,\wp(z)^{[(N-3)/2]}\right)\wp'(z) = \frac{dx}{
w}\left(1,\dots,x^{[(N-3)/2]}\right), \end{equation} where the
coefficients of the expansion are such as to cancel all poles
except for the two simple poles of (\ref{sj}). The differential
(\ref{u})  and the differentials (\ref{sj}) are single-valued on
the torus. In contrast to this (\ref{stau}) is a {\em
multi-valued} differential. Before turning to the residue formulae
we first explain how the Riemann bilinear identity works for these
latter differentials.

The general theory gives (compare with \cite{KriW})
\begin{align}\label{fu}\frac{\d\F}{ \d\a}
&=\frac{\d\F}{ \d u}=-\tau \log w_0- \oint_{B} dS=-\int_{z_0}\sp{z_0+\tau}
\log w\, dz, \\
\label{fak}\frac{\d\F}{ \d a_k} &= \int_{P_k}^{P_N} {\tilde z} dz, \\
\frac{\d\F}{\d\tau} &= -
 \oint_A {\tilde z}dS=\frac12\left(\log w_0\right)\sp2-\frac12
 \oint_A \left(\log w\right)\sp2dz.
 \label{ft}
\end{align}
As a consequence we obtain
\begin{align}\label{ftu}\frac{\d\sp2\F}{\d\tau \d u}&=- \log w_0- \oint_{B}
d\Omega_\tau, \\
\frac{\d\sp2\F}{\d u\d\tau} &= -
 \oint_A \log w\,dz=- \log w_0+ \oint_A z\frac{dw}{w}
 \label{fut}
\end{align}
and to show the integrability of $\cal{F}$ we must further
investigate the  {\em multi-valued} differential (\ref{stau}).
Using the fact that theta functions satisfy the heat equation we
may write
\begin{align*} 
d\Omega_\tau &= dz\left(\sum_{j=1}^{N}\d_\tau\log \theta_1(z-a_j) -
N\d_\tau\log \theta_1(z)\right) \\
&=\frac{dz}{ 4\pi i}\left(\sum_{j=1}^{N}(\log \theta_1(z-a_j))'' -
N(\log
\theta_1(z))''\right) 
+ \frac{dz}{ 4\pi i}\left(\sum_{j=1}^{N}\left((\log
\theta_1(z-a_j))'\right)^2 - N\left((\log
\theta_1(z))'\right)^2\right). \nonumber
\end{align*}
Here prime means derivative with respect to $z$ and in the final
equality we have separated the single-valued part of
$d\Omega_\tau$ (the first term) from its  multi-valued part. This
gives rise to
\begin{align} \nonumber
\Delta_B d\Omega_\tau &=
d\Omega_\tau (z+\tau) - d\Omega_\tau(z) = \frac{dz}{ 4\pi
i}\left.\left(\sum_{j=1}^{N}\left((\log \theta_1(z-a_j))'\right)^2
-
N\left((\log \theta_1(z))'\right)^2\right)\right|_z^{z+\tau} \\
&= -dz\left(\sum_{j=1}^{N}(\log \theta_1(z-a_j))' - N(\log
\theta_1(z))'\right) = -\frac{dw}{ w} .\label{otjump}
\end{align}
In more invariant terms \cite{KriW} this can be stated as the
``jump"  of the non-single valued differential (\ref{stau}) across
the $A$-cycle, with  jump $\Delta_B d\Omega_\tau =d\Omega_\tau^+ -
d\Omega_\tau^- = -{dw}/{ w}$. From (\ref{stau}) and the $\tau$
independence of (\ref{auw}) it also follows that
\begin{equation}\label{padw}0=\oint_A d\Omega_\tau.\end{equation}
The integrability condition for (\ref{ftu},\ref{fut}) now follows
upon considering the integral over the boundary $\d\Sigma$ of the
cut $z$-torus
\begin{align}
 0 &= \int_{\d\Sigma} zd\Omega_\tau =
\oint_A\left(zd\Omega_\tau - (z+\tau)(d\Omega_\tau -\frac{dw}{
w})\right) + \oint_B\left((z+1)d\Omega_\tau - zd\Omega_\tau\right) =
\oint_A z\frac{dw}{ w} + \oint_B d\Omega_\tau.
\label{rbi2}\end{align} Thus $ {\d^2 \F}/{\d\tau\d u} = {\d^2\F}/{\d
u\d\tau} $. The equality (\ref{rbi2}) also follows upon
differentiating
$$
0 = \int_{\d\Sigma} z^2\frac{dw}{w} = \oint_A\left(z^2-
(z+\tau)^2\right)\frac{dw}{w} +
\oint_B\left((z+1)^2-z^2\right)\frac{dw}{w}= 2\tau \oint_A
dS-2\oint_B dS,$$ which also establishes that
\begin{equation}
\label{costr} \tau\, \a=\oint_B dS.
\end{equation}
This then leads to a term $\ha\tau u^2$ in the prepotential.

Let us now turn to the residue formula (\ref{residue}). The
addition of the extra variables now mean further terms to those
calculated in \cite{MaMi},
\begin{align} \label{fijk} \frac{{\d}\sp3 \mathcal{F}}{ \d a_j\d a_j\d a_k} &=
\ \res_{\frac{dw}{ w}=0} \left(\frac{d\Omega_id\Omega_jd\Omega_k}{
dz \frac{dw}{ w}}\right)
\\
&= \left\{ \begin{array}{c}
  2\sum_{l\ne N}\hat\zeta (a_{lN})-
\sum_{l\ne i,j,k,N}\hat\zeta (a_{lN})+N\hat\zeta (a_{N}),\qquad
\qquad
\qquad \ \ i\neq j\neq k, \\ \\
  -\hat\zeta (a_{ik})+
4\hat\zeta (a_{iN})+2\hat\zeta (a_{kN})+\sum_{l\ne i,k,N}\hat\zeta
(a_{lN})+N\hat\zeta (a_{N}),\
\ \ \ i=j\neq k, \\ \\
  \sum_{l\ne i}\hat\zeta( a_{il})+
6\hat\zeta (a_{iN})+\sum_{l\ne N}\hat\zeta
(a_{lN})+N\left(\hat\zeta (a_{N})-\hat\zeta (a_{i})\right),\ \ \ \
i=j=k.
\end{array}
\right. \end{align} Here $a_{ij} = a_i-a_j$ and $\hat\zeta(z)
\equiv \frac{d}{ dz}\log\theta_1(z)$. (The function $\hat\zeta(z)$
differs from the usual Weierstrass $\zeta$-function by a term
linear in $z$: $\hat\zeta(z)=\zeta(z)-2\eta z$.) Additionally we
have
\begin{align} \label{fuut}
\frac{\d^3 \F}{ \d u\d u\d\tau} &= \ \res_{\frac{dw}{
w}=0}\left(\frac{dz dz d\Omega_\tau}{ dz \frac{dw}{ w}}\right) = \
\res_{\frac{dw}{ w}=0}\left(\frac{dz d\Omega_\tau}{\frac{dw}{
w}}\right)
 = \oint _{\d\Sigma_{\rm cut}} \left(\frac{dz d\Omega_\tau}{
\frac{dw}{ w}}\right) \nonumber  \\&= \oint _{A} dz
\frac{d\Omega_\tau^--d\Omega_\tau^+}{ \frac{dw}{ w}} = \oint
_{A}dz =1,\\
\label{fuuu} \frac{\d^3 \F}{ \d u^3} &= \ \res_{ \frac{dw}{
w}=0}\left( \frac{dz dz dz}{ dz  \frac{dw}{ w}}\right) = \ \res_{
\frac{dw}{ w}=0}\left( \frac{dz dz}{ \frac{dw}{ w}}\right) = 0,\\
 \label{futt}
 \frac{\d^3 \F}{ \d u\d \tau\d\tau} &=  \res_{ \frac{dw}{ w}=0} \left( \frac{dz
d\Omega_\tau d\Omega_\tau}{ dz  \frac{dw}{ w}}\right) =
\res_{\frac{dw}{ w}=0}
\left(\frac{d\Omega_\tau d\Omega_\tau}{\frac {dw}{ w}}\right) \nonumber \\
&= \oint _{A} \frac{(d\Omega_\tau^+)^2-(d\Omega_\tau^-)^2}{
\frac{dw}{ w}} = -\oint _{A}\left( d\Omega_\tau^+ +
d\Omega_\tau^-\right) = 0. \end{align} Here we have used
(\ref{otjump})  and (\ref{padw}). Together these means that the
tri-logarithmic expression of \cite{MaMi} should be corrected by
adding the term $\half \tau u^2$, consistent with \rf{costr}, and
some function of $\tau$, computed in \cite{str}, which can be fixed by
the residue formula for ${\d^3 \F}/{ \d u\sp3}$.

\subsection{WDVV for the perturbative 6d prepotential}

From the formulas of the previous section it is obvious that the
6d perturbative prepotential satisfies the WDVV
equations (\ref{WDVV}) as a function of the $n_m=N+1$ parameters
$\{ T_I \} = \{u, a_j\ (j=1\dots N-1),\tau\} $. Indeed, having the
residue formulas, one has only to check the matching condition
(\ref{matching}). If this holds the WDVV equations then simply
follow from the associativity of the algebra of functions at the
critical points $\{ z_\alpha\}$, the solutions of
\begin{equation} \label{clpo}
\frac{dw}{ w} =
\left(\sum_{j=1}^{N}\frac{\theta_1'(z-a_j)}{\theta_1(z-a_j)} -
N\frac{\theta_1'(z)}{\theta_1(z)}\right)dz = 0. \end{equation} Now
since the differential (\ref{clpo}) obviously has $N+1$ poles then
it also has $\# (\alpha) =N+1$ zeroes. Thus $n_z = N+1 = n_m$ and
the matching condition holds.

The corresponding associative algebra is simply realized as the
algebra of functions at the points $\{ z_\alpha\}$, with any
appropriate choice of $\xi$. The corresponding basis can be chosen
as \begin{equation} \label{fio}
\begin{split}
\phi_j &= \frac{d\Omega_i}{ dz} =
\frac{\theta_1'(z-a_{N})}{\theta_1(z-a_{N})}
-\frac{\theta_1'(z-a_j)}{\theta_1(z-a_j)},\ \ \ j=1,\dots,N-1,
\\
\phi_\tau &= \frac{d\Omega_\tau}{ dz} =
\sum_{j=1}^{N}\frac{\theta_1''(z-a_j)}{\theta_1(z-a_j)}
-N\frac{\theta_1''(z)}{\theta_1(z)},
\\
\phi_z &= 1. \end{split} \end{equation} Our algebra of functions
here `accidentally' contains the ``unity" $\phi_z=1$, but we
stress that this does not influence any of our statements made
about the WDVV equations beyond the specific $u$-dependence of the
prepotential  via the $\half \tau u^2$-term. This simple
dependence almost certainly does not survive beyond the
perturbative limit of the 6d theory.

The only delicate point to note here is that $\phi_\tau$
(\ref{fio}) is {\em not} single valued on the torus $\Sigma$.
However, this is not a problem when considering the associative
algebra (\ref{eqc}) since we restrict the values of the functions
(\ref{fio}) to their {\em values} at critical points $\{
z_\alpha\}$, where the ambiguity in the definition of $\phi_\tau$
disappears,
\begin{equation} \label{faj} \left.\left(\phi_\tau^+ -
\phi_\tau^-\right)\right|_{z=z_\alpha} =
\left.\left(\frac{d\Omega_\tau^+}{ dz} - \frac{d\Omega_\tau^-}{
dz}\right)\right|_{z=z_\alpha} = \left.\left(\frac{d\log w}{
dz}\right)\right|_{z=z_\alpha} = 0. \end{equation} Therefore the
quantities $\phi_\tau(z_\alpha)$ are uniquely defined.

For later comparison it is instructive to write down the simplest
$SU(2)$ case with $N+1=3$, when the WDVV equations (\ref{WDVV})
are already nontrivial. In this case (upon noting
$\sigma(z)=e\sp{\eta z^2}\theta_1(z)$ and
 $\wp(z)=-\zeta'(z)=-2\eta- \frac{d^2}{ dz^2}\log\theta_1(z)$)
\begin{equation} \label{n1} w = e^u
\frac{\theta_1(z-a)\theta_1(z+a)}{ \theta_1(z)^2} =
\theta_1\sp2(a)e^u\left(\wp(a)-\wp(z)\right) \end{equation} and
\begin{equation}
\frac{dw}{ w} = \frac{\wp'(z) dz}{ \wp(z) - \wp(a)}.
\end{equation} The latter has three poles at $z=0$, $\pm a$ and three
zeroes at the half-periods of the $z$-torus
\begin{equation} \label{hape} \{ z_\alpha \} = \{ \omega_\alpha \} = \ha,\
\frac{\tau}{ 2},\ \frac{1+\tau}{ 2}. \end{equation} The basis of
functions (\ref{fio}) for this case is
\begin{align}
\label{fio2} \phi_a &= \frac{d\Omega_a}{ dz} =
\frac{\theta_1'(z+a)}{\theta_1(z+a)}
-\frac{\theta_1'(z-a)}{\theta_1(z-a)} =
\hat\zeta(z+a)-\hat\zeta(z-a) = 2\hat\zeta(a) -
\frac{\wp'(a)}{\wp(z)-\wp(a)},
\\
\phi_\tau &= \frac{d\Omega_\tau}{ dz} =
\frac{\theta_1''(z+a)}{\theta_1(z+a)}
+\frac{\theta_1''(z-a)}{\theta_1(z-a)}-2\frac{\theta_1''(z)}{\theta_1(z)},
\\
\phi_z &= 1. \end{align} Observe that the curve (\ref{n1}) in the
$SU(2)$ case has an additional symmetry $z \leftrightarrow -z$. We
will see below that the WDVV equations hold generally for a
subfamily of bi-elliptic curves with this extra symmetry.

\subsection{The non-perturbative 6d theory}

Consider now the non-perturbative 6d $SU(N)$ theory
associated to the curve
\begin{equation}
\label{npcu} w+\frac{\Lambda^{2N}}{ w} =
e^u\,\prod_{j=1}^{N}\frac{\theta_1(z-a_j)}{ \theta_1(z)}\equiv
e^u\,H(z),\qquad \sum_{j=1}^{N}a_j=0,
\end{equation} and generating differential $dS = -z
\frac{dw}{ w}$. These curves have the $\mathbb{Z}_2$ symmetry
$\sigma:w\leftrightarrow{\Lambda^{2N}/ w}$ under which $dS$ is
odd, and when $N=2$ there is the additional symmetry $z
\leftrightarrow -z$ noted above. The curves have genus $g=N+1$ for
the $SU(N)$ gauge theory. We may picture this curve as two
$z$-tori glued along $N$ identical cuts on each. The ends of these
are located at the zeroes of $H(z)^2-4$. One may choose a
canonical basis of cycles with $A_1$ and $A_{N+1}$ being
$A$-cycles on the $z$-tori while the $A_j$-cycle ($j=2,\dots,N$)
surrounds the $(j-1)$-st handle joining them. (These handles
degenerate to pairs of points in the perturbative limit.) The
corresponding $B$-cycles are determined by $A_i\circ
B_j=\delta_{ij}$. The cycle $B_j$ may be taken as going from one
torus through the $(j-1)$-st handle and returning through the
$N$-th handle. This choice of cycles can be made so that
$\sigma_\star(A_1)=A_{N+1}$, $\sigma_\star(B_1)=B_{N+1}$ and
$\sigma_\star(A_j)=-A_{j}$, $\sigma_\star(B_j)=-B_j$
($j=2,\dots,N$).

For illustrative purposes we focus first on the $SU(2)$ case with
the generalization to arbitrary $N$ being briefly given later in
\begin{equation} \label{2npc} w+\frac{\Lambda\sp4}{ w}
=e^{u}H(z)=e^{u}\theta_1\sp2(a)\left(\wp(a)-\wp(z)\right).
\end{equation}
Now we have two $z$-tori glued along two identical cuts, the ends
which are located at the zeroes of the section. The $SU(2)$ case
gives us
\begin{align} \label{npy} \left(w-\frac{\Lambda\sp4}{ w}\right)^2
&=e^{2u} H(z)^2-4\Lambda\sp4 \nonumber\\
&=e^{2u}\theta_1\sp4(a)\left(\wp(z)-\wp(a)-2\tilde\Lambda\sp2\right)
\left(\wp(z)-\wp(a)+2\tilde\Lambda\sp2\right), \
\tilde\Lambda\sp2=\frac{\Lambda\sp2}{e\sp{u}\theta_1\sp2(a)}.
\end{align}
We may choose as a basis of holomorphic differentials
\begin{equation} \label{hol3} dz, \ \ \ \ \frac{w+\frac{\Lambda\sp4}{ w}}{
w-\frac{\Lambda\sp4}{w}}\left(\frac{\theta_1'(z+a)}{\theta_1(z+a)}
-\frac{\theta_1'(z-a)}{\theta_1(z-a)}\right)dz,\ \ \ \
\frac{w+\frac{\Lambda\sp4}{ w}}{ w-\frac{\Lambda\sp4}{w}}\,dz,
\end{equation} {or equivalently
\begin{equation} \label{hol3'} dz, \ \ \ \
\frac{dz}{ w-\frac{\Lambda\sp4}{ w}},\ \ \ \
\frac{w+\frac{\Lambda\sp4}{ w}}{ w-\frac{\Lambda\sp4}{w}}\,dz,
\end{equation}the two being related by
\begin{equation} \label{licom}
\frac{w+\frac{\Lambda\sp4}{ w}}{ w-\frac{\Lambda\sp4}{
w}}\left(\frac{\theta_1'(z+a)}{\theta_1(z+a)}
-\frac{\theta_1'(z-a)}{\theta_1(z-a)}\right)dz = 2\hat\zeta(a)
\frac{w+\frac{\Lambda\sp4}{ w}}{ w-\frac{\Lambda\sp4}{ w}}dz + e^u
\theta_1\sp2(a)\wp'(a)\frac{dz}{ w-\frac{\Lambda\sp4}{ w}}.
\end{equation}

Now consider the variation of the generating differential $dS$ (at
constant $w$). Up to total differentials \begin{equation}
\begin{split} \label{vardS} \delta\left(-z\frac{dw}{ w}\right) &=
\delta u\ dz\frac{w+\frac{\Lambda\sp4}{w}}{
w-\frac{\Lambda\sp4}{w}}  + \delta a\
\frac{w+\frac{\Lambda\sp4}{w}}{w-\frac{\Lambda\sp4}{w}}\left(\frac{\theta_1'(z+a)}{\theta_1(z+a)}
-\frac{\theta_1'(z-a)}{\theta_1(z-a)}\right)dz \\
&\qquad+ \frac{\delta\tau}{4\pi i} \
\frac{w+\frac{\Lambda\sp4}{w}}{w-\frac{\Lambda\sp4}{w}}\left(\frac{\theta_1''(z+a)}{\theta_1(z+a)}
+\frac{\theta_1''(z-a)}{\theta_1(z-a)}-2\frac{\theta_1''(z)}{\theta_1(z)}\right)dz.
\end{split}
\end{equation} The first two terms of this expansion are holomorphic Abelian differentials
with the final term, corresponding to $\delta\tau$, being non-single
valued. This is just as in the perturbative case. Note that the
holomorphic differential $dz$ does not appear in the expansion of
$\delta (dS)$. Indeed this has a different behaviour under the
$\mathbb{Z}_2$ symmetry (now $w\leftrightarrow\frac{\Lambda\sp4}{
w}$) compared to the generating differential $dS$: $dS$ itself and
all the constituents of (\ref{vardS}) are {\em odd} with respect to
this symmetry, while $dz$ is even. This latter differential is
related with the deformation of $dS$ with respect to the scale
parameter
$$\d_{\log\Lambda\sp2}\, dS\big|_{\frac{w}{\Lambda\sp2}}=dz.$$

Suppose now that $\{d\omega_i\}$ are canonical holomorphic
differentials such that $\oint_{A_i}d\omega_j=\delta_{ij}$. Then
using the symmetry under $\sigma$ we  have expansions
\begin{align}
\label{cano} d\omega_{-}&\equiv d\omega_{1}-d\omega_{3} =
\alpha_-\frac{w+\frac{\Lambda\sp4}{w}}{w-\frac{\Lambda\sp4}{w}}\
dz +
\beta_-\frac{w+\frac{\Lambda\sp4}{w}}{w-\frac{\Lambda\sp4}{w}}
\left(\frac{\theta_1'(z+a)}{\theta_1(z+a)}
-\frac{\theta_1'(z-a)}{\theta_1(z-a)}\right)dz,
\\
d\omega_2& = \alpha
\frac{w+\frac{\Lambda\sp4}{w}}{w-\frac{\Lambda\sp4}{w}}\ dz
+\beta\
\frac{w+\frac{\Lambda\sp4}{w}}{w-\frac{\Lambda\sp4}{w}}\left(\frac{\theta_1'(z+a)}{\theta_1(z+a)}
-\frac{\theta_1'(z-a)}{\theta_1(z-a)}\right)dz,
\end{align} where the coefficients $\alpha_-$, $\beta_-$,
$\alpha$, $\beta$ are determined from \begin{align} \label{nA1}
\oint_{A_2}d\omega_- &= 0, \qquad\oint_{A_1}d\omega_- = 1=
-\oint_{A_3} d\omega_- ,\qquad  \oint_{A_2}d\omega_2 = 1,
\qquad\oint_{A_1}d\omega_2 = 0= -\oint_{A_3} d\omega_2 .
\end{align}
In the perturbative limit $\frac{w+\frac{\Lambda\sp4}{w}}{
w-\frac{\Lambda\sp4}{w}}\to \pm 1$, depending on which base torus
solution we choose to (\ref{2npc}). Similarly
$\frac{w+\frac{\Lambda\sp4}{w}}{ w-\frac{\Lambda\sp4}{w}}\ dz\to \pm
dz$. The differential
$\frac{w+\frac{\Lambda\sp4}{w}}{w-\frac{\Lambda\sp4}{w}}\left(\frac{\theta_1'(z+a)}{\theta_1(z+a)}
-\frac{\theta_1'(z-a)}{\theta_1(z-a)}\right)dz$ becomes an Abelian
differential of the third kind with poles at $z=\pm a$, the two cuts
reducing to these two points. In this limit we have
\begin{equation} \label{pcoef} \left(
\begin{array}{cc}
  \alpha_- & \beta_- \\
  \alpha & \beta
\end{array}\right) \rightarrow
\left( \begin{array}{cc}
  1 & 0 \\
  0 & 1
\end{array}\right),
\end{equation} consistent with (\ref{nA1}),  and consequently
\begin{equation} \label{B2z} \oint_{B_2} d\omega_- \rightarrow
2\int_{-a}^a dz. \end{equation}  In this way we recover our
earlier perturbative results.

It remains to describe the matching conditions necessary for the
WDVV equations. We see from (\ref{vardS}) that we have $n_m=3$
moduli here. On the nonperturbative curve (\ref{npcu}) a
holomorphic differential has $2(g-1)=4$ zeroes, while the
differential $\frac{dw}{ w}$ has two simple poles and therefore
$2+2(g-1)=6$ zeroes. However these six zeros arise from $n_z=3$
values $\{z_\alpha\}$ (with two $w$ values for each). For the case
at hand the half-periods (\ref{hape}) $z_\alpha=\omega_\alpha$,
$\alpha=1,\dots,3$, are solutions to $H'(z)=0$. The differentials
in the residue formula only depend on $z_\alpha$ here, and so the
symmetry of the curve enables us to get matching here.

Let us now consider the general $N$ case which goes through in
much the same way. Instead of (\ref{vardS}) one now has (upon
setting $\Lambda=1$ hereafter)
\begin{equation}
\begin{split}
\label{vardSN} \delta\left(-z\frac{dw}{w}\right) &= \delta u\
dz\frac{w+\frac{1}{w}}{w-\frac{1}{w}} + \sum_{j=1}^{N-1}\delta
a_j\
\frac{w+\frac{1}{w}}{w-\frac{1}{w}}\left(\frac{\theta_1'(z-a_N)}{\theta_1(z-a_N)}
-\frac{w+\frac{1}{w}}{
w-\frac{1}{w}}\frac{\theta_1'(z-a_j)}{\theta_1(z-a_j)}\right)dz
\\
&\qquad+\frac{\delta\tau}{4\pi i} \
\frac{w+\frac{1}{w}}{w-\frac{1}{w}}\left(\sum_{j=1}^{N}
\frac{\theta_1''(z-a_j)}{\theta_1(z-a_j)}-N\frac{\theta_1''(z)}{\theta_1(z)}\right)dz.
\end{split}
\end{equation} The right hand side of
(\ref{vardSN}) now consists of the $N$ (out of the total $N+1$)
holomorphic differentials that are odd with respect to the
$\mathbb{Z}_2$ symmetry $w\leftrightarrow\frac{1}{w}$. In analogy
with (\ref{hydeg}) these may be expanded over the non-perturbative
basis
\begin{align} \label{hye} \frac{dz}{ w-\frac{1}{w}}\left(1,\dots,\wp(z)^{[N/2]}\right) &=
\frac{dx}{ \left(w-\frac{1}{w}\right)y}\left(1,\dots,x^{[N/2]}\right),\\
 \label{hyo} \frac{dz}{ w-\frac{1}{w}}\left(1,\dots,\wp(z)^{[(N-3)/2]}\right)\wp'(z) &=
 \frac{dx}{
w-\frac{1}{w}}\left(1,\dots,x^{[(N-3)/2]}\right). \end{align} Now
all of the differentials appearing in the residue formula are
$\mathbb{Z}_2$-odd. Upon writing
$d\Omega_I=h_I(z)\frac{w+\frac{1}{w}}{w-\frac{1}{w}}dz$ the
residue formula (\ref{residue}) for the prepotential $\F$ becomes
\begin{equation}\label{resnp}
\begin{split}
\F_{IJK} &= \res_{\frac{dw}{w}=0}\
\left(\frac{w+\frac{1}{w}}{w-\frac{1}{w}}\right)^3
\frac{h_I(z)h_J(z)h_K(z) dz^3}{ dz\frac{dw}{w}} =
\sum_\alpha\res_{z_\alpha} \frac{H(z)^3h_I(z)h_J(z)h_K(z) dz}{
(H(z)^2-4)H'(z)}
\\
&=
\sum_\alpha\frac{H(z_\alpha)^3h_I(z_\alpha)h_J(z_\alpha)h_K(z_\alpha)
}{ (H(z_\alpha)^2-4)H''(z_\alpha)}\end{split}
\end{equation} where
$z_\alpha$ are solutions of $dw/w=0$. To describe what these are,
first observe that $dz$ and $\frac{dz}{w-\frac{1}{w}}$ are
holomorphic differentials and so have $2(g_\Sigma-1)=2N$ zeros. The
divisor of $dz$ are the $2N$ branch points of
$$\left(w-\frac{1}{ w}\right)^2
=e^{2u}\left( H(z)^2-4\right),$$ the points at which $w-\frac{1}{
w}=0$. The divisor of $\frac{dz}{w-\frac{1}{w}}$ are the points
$z=0$ and $w=0,\infty$. Now $H'(z)$ has a pole of order $N+1$ at
$z=0$. This pole cancels the zeros of $\frac{dz}{w-\frac{1}{w}}$
leaving a simple pole remaining (one at $(z=0,w=0)$ and one at
$(z=0,w=\infty)$) and the zeros of
$$\frac{dw}{w}=e\sp{u}H'(z)\,\frac{dz}{w-\frac{1}{w}}$$
are the $(N+1)$ (as a function of $z$) zeros $\{z_\alpha\}$ of
$H'(z)$, and so corresponds to $2(N+1)$ points on the curve
$\Sigma$. Here we have $n_m=N+1$ moduli and, because of the symmetry
of the differentials $d\Omega_I$, $n_z=\#\{z_\alpha\}=N+1$. This
means that the non-perturbative $\F$, defined by (\ref{resnp}),
satisfies the WDVV equations as a function
$\F(T_K)=\F(\a_1-\a_{N+1},\a_2,\dots,\a_N,\tau)$ of $N$ linear
combinations of the $N+1$ SW periods $\a_i = \oint_{A_i}dS$ and the
modular parameter $\tau$ of the base curve.

\section{A bi-elliptic generalization}

We now consider generalizing our discussion to the setting when the
variable $\tilde z$ is also elliptic. First observe that the
non-perturbative $SU(2)$ curve (\ref{2npc}), (\ref{npy}) can be
rewritten in ``hyperelliptic" terms as
\begin{equation} \label{nphe} y^2 = \prod_{i=1}^3(x-e_i),\ \ \ \ Y^2
= (x-b_+)(x-b_-),
\end{equation}
where $x=\wp(z)$ and $y=-\half\wp'(z)$ are the affine coordinates of
the torus (\ref{tauto}) and
$Y=e^{-u}\left(w-\frac{1}{w}\right)/\theta_1\sp{2}(a)$. By replacing
the second equation in (\ref{nphe}) with a polynomial of fourth
degree on the right hand side,
\begin{equation} \label{twoto2} y^2 = \prod_{i=1}^3(x-e_i), \ \ \ \
Y^2 = \prod_{j=1}^4(x-b_j),
\end{equation} one comes to a sort of a ``bi-elliptic" system that
we will further explore. (We reserve the term ``double-elliptic"
for the system considered in \cite{dell}.) Equivalently via
the fractional-linear transformation
\begin{equation} {\tilde x} = \frac{ax+b}{ cx+d}, \ \ \ \ \ x = \frac{d{\tilde
x}-b}{ -c{\tilde x}+a}, \ \ \ \ \ Y \propto (cx+d)^2{\tilde y},
\end{equation}the
curve (\ref{twoto2}) may be rewritten as two Weierstrass equations
\begin{equation} \label{twoel} y^2 = \prod_{i=1}^3(x-e_i),\ \ \ \
{\tilde y}^2 = \prod_{i=1}^3({\tilde x}-{\tilde e}_i).
\end{equation}
We remark that the curve (\ref{twoto2}) has the symmetry
$\mathbb{Z}_2\times\mathbb{Z}_2$: $z\leftrightarrow -z$ and ${\tilde
z}\leftrightarrow -{\tilde z}$, where now $x=\wp(z;\tau)$,
$y=-\frac12\wp'(z;\tau)$ and $\tilde x=\wp(\tilde z;\tilde \tau)$,
$\tilde y=-\frac12\wp'(\tilde z;\tilde \tau)$ are the standard
coordinates of the tori.

The genus of the curve (\ref{twoto2}) is $g=5$. This coincides with
the total number of holomorphic differentials, which are linear
combinations of
\begin{equation} \label{hdtt} \frac{dx}{ yY},\ \ \ x\frac{dx}{ yY},\ \
\ x^2\frac{dx}{ yY}, \ \ \ \frac{dx}{ y}=-2dz,\ \ \ \frac{dx}{
Y}=\frac{d{\tilde x}}{ {\tilde y}}=-2d{\tilde z}. \end{equation}
Taking into account the $\mathbb{Z}_2\times\mathbb{Z}_2$ symmetry
the number of zeroes of the holomorphic differentials (in
particular of $dz$ and $d{\tilde z}$) is seen to be $2(g-1) = 8$.
Equally the genus may calculated using the Riemann-Hurwitz
formula. The curve may be viewed as a 4-sheeted cover of the
$x$-plane, with each sheet corresponding to particular choice of
the sign for $(y,Y)=(\pm,\pm)$. There are a total of $B=2\cdot 4 +
2\cdot 4 = 16$
 branch points and so. Then \begin{equation} \label{rh}
g-1 = \#\ ({\rm sheets})(g_0-1)+B/2 = 4(0-1) +16/2 =4,
\end{equation} or $g=5$.

The example of (\ref{twoto2}) may be extended as follows. We
consider the $\mathbb{Z}_2\times\mathbb{Z}_2$ symmetric curve
arising when the two elliptic equations (\ref{twoel}) are
rationally related via
\begin{equation}
\label{ratmap} \tilde x = \frac{P_m(x)}{ R_m(x)} =
\frac{p_mx^m+\dots+p_0}{ r_mx^m+\dots+r_0}.
\end{equation}
Upon introducing $Y \propto R_m(x)^2\tilde y$ this may be
rewritten in a form similar to (\ref{twoto2}),
\begin{equation}
\label{twoto}
y^2 = \prod_{i=1}^3(x-e_i),\ \ \ \ Y^2 =
R_m(x)\prod_{j=1}^3(P_m(x)-{\tilde e}_jR_m(x))
\end{equation}
where on the right hand side of the last formula we now have a
polynomial of degree $4m$. The trigonometric and rational
degenerations of (\ref{ratmap}) then take the form
\begin{equation}
\label{trim} w+\frac{1}{w} = P_m(x)
\end{equation}
and
\begin{equation}
\label{ram} w = P_m(x)
\end{equation}
respectively. In contrast to (\ref{6dc}) the righthand side here
depends only on the Weierstrass function $x=\wp(z)$ which means that
in this symmetric case one considers only the even differentials
(\ref{hye}) and of (\ref{hydeg}) with $[N/2]=m$.

The genus of the curve (\ref{twoto}) is $g=4m+1$. For a fixed $x$
there are $4m$ $Y$-branch points, and as there are two $z$'s for
each $x$ we obtain $8m$ branch points. The total number of branch
points is then $B=2\cdot 4 + 2\cdot 4m$ and application of the
Riemann-Hurwitz formula (\ref{rh}) gives the stated genus. It is
also clear that the curve defined by (\ref{ratmap}) together with
(\ref{twoel}), or by (\ref{twoto}), can be visualized as a
double-cover of the $z$-torus with $8m$ branch points. Taking the
derivative of (\ref{ratmap}), one gets \begin{equation}
\label{tytz} \tilde y d\tilde z = \frac{P_m'(x)R_m(x) -
P_m(x)R_m'(x)}{ R_m(x)^2} ydz
\end{equation} or
\begin{equation}
\label{zdtz} Y d\tilde z = (P_m'(x)R_m(x) - P_m(x)R_m'(x)) ydz.
\end{equation}
The number of zeroes of $d{\tilde z}$ is given by zeroes of the
righthand side of (\ref{zdtz}) and equals
\begin{equation}
\label{nzdtz} 4\cdot (2m-2)+2\cdot 4 = 8m = 2(g-1).
\end{equation}
Here the first factor of $4$ comes from taking into account the
total $\mathbb{Z}_2\times\mathbb{Z}_2$ multiplicity of the $2m-2$
zeros of the polynomial $P_m'R_m - P_mR_m'\sim x^{2m-2}$, while
the second term on the left-hand side of (\ref{nzdtz}) counts the
$4$ half-periods of the $z$-torus (including $z=0$ if compared to
(\ref{hape})), with the factor $2$ corresponding to the ${\tilde
z} \leftrightarrow - {\tilde z}$ part of
$\mathbb{Z}_2\times\mathbb{Z}_2$.

An alternative approach to obtain a generalization to that
outlined above would be to consider the case with the elliptic
cosine replacing the Weierstrass function $\tilde x$ in the
left-hand side of (\ref{ratmap}). Naively, this would lead to an
equation
\begin{equation} \label{ratmapco} \frac{\tilde x-\tilde e_1}{\tilde
x-\tilde e_2} = \frac{P_m(x)^2}{ R_m(x)^2} \end{equation} with the
polynomials on the righthand side being the squares of those
appearing in  (\ref{ratmap}). Analogous to our rewriting of
(\ref{twoto}) one now finds that
\begin{equation}
\label{twotoco} y^2 = \prod_{i=1}^3(x-e_i),\ \ \ \ Y^2 =
\left(P_m(x)^2-R_m(x)^2\right) \left({\tilde
e}_{23}P_m(x)^2-{\tilde e}_{13}R_m(x)^2\right),
\end{equation}
where now \begin{equation} \label{Yco} Y = {\tilde
y}\frac{P_m^2-R_m^2}{ P_mR_m{\tilde e}_{12}} \end{equation}
 and
${\tilde e}_{ij} \equiv {\tilde e}_i-{\tilde e}_j$. Again we find
a polynomial of degree $4m$ on the righthand side of
(\ref{twotoco}), and this case completely repeats our discussion
of the curve (\ref{twoto}).

One can also generalize to less symmetric curves where we only
have a $\mathbb{Z}_2$ symmetry with no extra $z\leftrightarrow -z$
or $y\leftrightarrow -y$ symmetry. The analysis of this parallels
what we have already presented and we will simply present the list
of results in the next section.

\section{Summary of general results}
We now present a summary of the results for the separated case of
the bi-elliptic curve (\ref{sepv}) when the function ${\cal H}$
may be expressed as a sum of two terms; they being elliptic
functions of $z$ and $\tilde z$. Let us start with the case of an
algebraic function ${\cal H}$ linear in the co-ordinate $\tilde x
= \wp(\tilde z|\tilde\tau)$,
\begin{equation} \Sigma_{\rm ell}:\ \ \ \tilde x = H_N(z) =
\frac{P(x) + yQ(x)}{R(x) + yS(x)} = e^u\prod_{i=1}^N
\frac{\theta(z-a_i|\tau)}{\theta(z-a'_i|\tau)}, \ \ \ \ \ dS=-z
d\tilde z, \label{ell} \end{equation} with polynomials $P(x)$ and
$R(x)$ of degree $\left[\frac{N}{2}\right]$ and the polynomials
$Q(x)$ and $S(x)$ of degree $\left[\frac{N-3}{2}\right]$. Further
$\sum_{i=1}^N (a_i - a'_i) = 0$. We shall call this the elliptic
case with $\mathbb{Z}_2$-symmetry; the
$\mathbb{Z}_2\times\mathbb{Z}_2$-symmetric subfamily
(\ref{ratmap}) corresponds to the vanishing of $Q$ and $S$ in the
righthand side of (\ref{ell}).

The non-perturbative 6d theory studied earlier corresponds to the
trigonometric degeneration of the $\tilde z$-torus in (\ref{ell}),
\begin{equation}
\Sigma_{\rm trig}:\ \ \ w+\frac{1}{w} = P(x) + yQ(x) =
\prod_{i=1}^N \frac{\theta(z-a_i|\tau)}{\theta(z|\tau)}, \ \ \
\sum_{i=1}^N a_i = 0, \ \ \ \ \ dS = -z\frac{dw}{w} .\label{trig}
\end{equation}
This may be viewed as an Inozemtsev limit (see \cite{Ino,IM2,SWbooks}),
when ${\tilde\tau} \rightarrow +i\infty$ and $\tilde z = i{\tilde
\tau}/2 - \log w$, so that $\tilde x \rightarrow {\rm const} +
2q^{1/4}(w+\frac{1}{w} + O(q^{9/4})$. After appropriate adjustment
of the polynomials the bi-elliptic family (\ref{ell}) then turns
then (\ref{trig}). This degeneration leads to the generic $SU(N)$
non-perturbative curve of genus $g=N+1$ discussed earlier. Upon
incorporating the scale $\Lambda$, redefining $w \rightarrow
\Lambda^{-2N}w$ and taking the perturbative limit $\Lambda
\rightarrow 0$, the system (\ref{trig}) further reduces to
(\ref{6dc}).

One observation to highlight from our analysis is that the number
of moduli appearing in the elliptic setting ($n_h$ in the tables
below) is roughly twice that of its trigonometric degenerations.

\subsection{Elliptic, trigonometric
and rational cases: the case of enlarged $\mathbb{Z}_2\times
\mathbb{Z}_2$ symmetry $\tilde z \leftrightarrow - \tilde z$ and
$z \leftrightarrow -z$}

\hspace{-1.5cm}
\begin{small}
\begin{tabular}{|c|c|c|c|}
\hline &&& \\
& elliptic (\ref{ratmap}) & trigonometric (\ref{trim}) &
rational (\ref{ram}) \\
&&&\\ \hline\hline &&&\\
spectral curve $\Sigma$& $\tilde x = \frac{P_m(x)}{R_m(x)}$ &
$w+\frac{1}{w} = P_m(x)$ & $w = P_m(x)$\\
&&&\\ \hline &&&\\
$n_m'$: &&&\\
$\#$ of parameters: the independent & $2(m+1)-1 = 2m+1$  & $m+1$ &
$m+1$\\
coefficients of polynomials &&&\\
&&&\\ \hline &&&\\
$n_m = n'_m + 1$ &&&\\
(inclusion of $\tau$) & $2m+2$ & $m+2$ & $m+2$ \\
&&&\\ \hline
&&&\\
${\cal H}'_{\tilde z}$ & 2$\tilde y$ &
$w-\frac{1}{w}$ & $w$ \\
&&&\\ \hline &&&\\
$Y^2 \sim ({\cal H}'_{\tilde z})^2$, & $Y^2=R^4_m(x) ({\cal
H}'_{\tilde z})^2 =$ &
$Y^2=\left(w-\frac{1}{w}\right)^2 =$ & $Y = w = P_m(x)$\\
expressed through $x$ & $R_m(x) \prod_{i=1}^3 \left(P_m(x)-\tilde
e_i R_m(x)\right)$&
$= P^2_m(x)-4$ &\\
&&&\\ \hline &&&\\
behaviour at large $x$ & $Y \sim x^{2m}$ & $Y \sim x^{m}$ &
$Y \sim x^{m}$ \\
&&&\\ \hline &&&\\
$\nu(\Sigma): \#$ of branch points of $\Sigma$&$8m$ & $4m$ & $2m$
punctures: the pairwise  \\
 over $z$-torus, where $Y=0$ &&&
``contracted" $4m$ branch points. \\
&&&\\ \hline &&&\\
genus $g(\Sigma) = \frac{\nu(\Sigma)-2}{2}+2$ &
$4m+1$&$2m+1$& $g=1$ torus with $2m$ punctures\\
&&&\\ \hline
\end{tabular}

\hspace{-1cm}
\begin{tabular}{|c|c|c|c|}
\hline &&&\\
&$\frac{(1,\ldots,x^{2m})dx}{Yy}$
&$\frac{(1,\ldots,x^{m})dx}{\left(w-\frac{1}{w}\right)y}$
& $\frac{(1,\ldots,x^{m})dx}{wy}$ \\
&&&\\
holomorphic differentials on $\Sigma$ &
$\frac{(1,\ldots,x^{2m-2})dx}{Y}$
&$\frac{(1,\ldots,x^{m-2})dx}{w-\frac{1}{w}}$
&  $\frac{(1,\ldots,x^{m-2})dx}{w}$\\
(or with simple poles&&&\\
 in rational case)&$\frac{dx}{y}=-2dz$ &$\frac{dx}{y}=-2dz$ &  $\frac{dx}{y}=-2dz$ \\
&&&\\ \hline &&&\\
total number of differentials& $(2m+1) + (2m-1)+1$ & $(m+1) +
(m-1) + 1 $ &
$(m+1) + (m-1) + 1 $\\
= $g(\Sigma)+\# {\rm marked\ points}$ & $=4m+1$ & $=2m+1$ & $=2m+1$\\
\hline
&&&\\
&$\mathbb{Z}_2\times \mathbb{Z}_2: \ \ $ & $\mathbb{Z}_2\times
\mathbb{Z}_2: \ \ $
& $\mathbb{Z}_2: ({\rm different}!) \ \ $ \\
symmetry of $\Sigma$ & $\left\{\begin{array}{cc}
z \leftrightarrow -z, & y\leftrightarrow -y\\
\tilde z \leftrightarrow -\tilde z,& Y \leftrightarrow -Y
\end{array}\right.$ &
$\left\{\begin{array}{cc}
z \leftrightarrow -z, & y\leftrightarrow -y\\
w \leftrightarrow \frac{1}{w},& Y \leftrightarrow -Y
\end{array}\right.$ &
$z \leftrightarrow -z$,\ $y\leftrightarrow -y$  \\
&&&\\ \hline &&&\\
holomorphic differentials, &$\frac{(1,\ldots,x^{2m})dx}{Yy}$&
$\frac{(1,\ldots,x^{m})dx}{\left(w-\frac{1}{w}\right)y}$ &
$\frac{(1,\ldots,x^{m})dx}{wy}$\\
{\it odd} under the symmetry
&&& $\frac{dx}{y}=-2dz$\\
&&&\\ \hline &&&\\
$n_h''$: $\#$ of such differentials & $2m+1$ &$m+1$ &$(m+1)+1=m+2$\\
&&&\\ \hline &&&\\
$n_h'$: $\#$ of hol. differentials, & $n_h' = n_h'' = 2m+1$ &
$n_h' = n_h'' = m+1$ &
$n_h' = n_h'' - 1 = m+1$\\
contributing to $\delta(d\tilde S)$ &&&$dz$ does not contribute\\
&&&\\ \hline &&&\\
$n_h = n'_h+1$ due to addition &&&\\
of $d\Omega_\tau = \frac{\partial (dS)}{\partial\tau}$
with jump & $2m+2$ & $m+2$ & $m+2$\\
&&&\\ \hline &&&\\
jump $\Delta_B (d\Omega_\tau)$ & $d\tilde z$&$\frac{dw}{w}$&
$\frac{dw}{w}$\\
&&&\\ \hline
\end{tabular}

\hspace{-1.5cm}
\begin{tabular}{|c|c|c|c|} \hline
&&&\\
 & & $2$ simple poles & $2m+1$ simple poles \\
$d\tilde z$ or $\frac{dw}{w}$ & holomorphic &at $x=\infty$
($z=0)$& at
$x=\infty$ ($z=0$)\\
&&related by $w \leftrightarrow \frac{1}{w}$& and zeroes of $P_m(x)$\\
&&&doubled by $z \leftrightarrow -z$\\ &&&\\ \hline &&&\\
&$2(2m-2)+4=4m$ zeros& $2(m-1)+3 = 2m+1$ zeros & $2(m-1)+3 =$ \\
zeros of $d\tilde z$ &
 of $d\left(\frac{P_m(x)}{R_m(x)}\right) = \frac{P_m'R_m-P_mR_m'}{R_m^2}ydz$&
 of $dP_m(x) = P'_m(x)ydz$&
 $=2m+1$ zeros of \\
(i.e. zeros of $\mathcal{H}'_z$) &produces two {\it different}
zeros&
produces two {\it different} zeros & $dP_m(x)= P'_m(x)ydz$\\
&related by $\tilde z \leftrightarrow -\tilde z$&
related by $w \leftrightarrow \frac{1}{w}$&\\
&&&\\ \hline &&&\\
$n'_z$: number zeros & $8m = 2\cdot 4m =2\left(g(\Sigma)-1\right)$
& $4m+2 = 2\cdot (2m+1)=$&
$2m+1=$\\
of $d\tilde z$ or $\frac{dw}{w}$&&$=2\left(g(\Sigma)-1\right) + \#
{\rm poles}$
& $\# {\rm poles}$ \\
&&&\\ \hline &&&\\
$n_z$: $\#$ of critical points &&&\\
in residue formula: & $\frac{1}{4}(2\cdot 2(2m-2)) +
\frac{1}{2}(2\cdot 4)$ & $\frac{1}{4}(2\cdot 2(m-1)) +
\frac{1}{2}(2\cdot 3)$&
$\frac{1}{2}2(m-1) + 4 $\\
($\#$ of zeros of $d\tilde z$  &$ = 2m+2$& $=m+2$&$= m+2$\\
over symmetry)&&&\\
&&&\\ \hline\hline &&&\\
matching $n_m = n_h = n_z$ & + & + & + \\
WDVV equations & & & \\
&&&\\ \hline
\end{tabular}
\end{small}
\vspace{1.5cm}
\noindent

This table completes our analysis for the most symmetric case. We see that the
WDVV equations \rf{WDVV} hold in all three (perturbative, non-perturbative and
bi-elliptic) cases.

\subsection{Summary for elliptic, trigonometric
and rational cases: the generic case of only $\mathbb{Z}_2$
symmetry $\tilde z \leftrightarrow - \tilde z$}

\begin{small}
\hspace{-1.5cm}
\begin{tabular}{|c|c|c|c|}
\hline &&& \\
& elliptic (\ref{ell}) & trigonometric (\ref{trig}) &
rational (\ref{6dc}) \\
&&&\\ \hline\hline &&&\\
spectral curve $\Sigma$& $\tilde x = \frac{P(x) + yQ(x)}{R(x) +
yS(x)}$ &
$w+\frac{1}{w} = P(x) + yQ(x)$ & $w = P(x) + yQ(x)$\\
&&&\\ \hline &&&\\
 &&&\\
$n_m'$: $\#$ of parameters: &
$2\left(\left(\left[\frac{N}{2}\right] +1\right) +
\left(\left[\frac{N-3}{2}\right] + 1\right)\right)-1=$  &
$\left(\left[\frac{N}{2}\right]+1\right) +
\left(\left[\frac{N-3}{2}\right]+1\right) $ &
$\left(\left[\frac{N}{2}\right]+1\right) +
\left(\left[\frac{N-3}{2}\right]+1\right) $\\
coefficients of polynomials & $ =2N -1$&$= N$&$=N$\\
&&&\\ \hline &&&\\
$n_m = n'_m + 1$ &&&\\
(inclusion of $\tau$) & $2N$ & $N+1$ & $N+1$ \\
&&&\\ \hline &&&\\
${\cal H}'_{\tilde z}$ & $2\tilde y$ &
$w-\frac{1}{w}$ & $w$ \\
&&&\\ \hline &&&\\
$Y^2 \sim ({\cal H}'_{\tilde z})^2$, & $Y^2=(R+yS)^4 ({\cal
H}'_{\tilde z})^2 =
(R+yS)\cdot$ & $Y^2=\left(w-\frac{1}{w}\right)^2 =$ & $Y = w = P+yQ$\\
expressed through $x$ & $\cdot \prod_{i=1}^3 \left((P-\tilde e_i
R) + y(Q-\tilde e_i S)\right)$&
$= (P+yQ)^2-4$ &\\
&&&\\ \hline &&&\\
behaviour at large $x$ & $Y \sim x^{N}$ &
$Y \sim x^{N/2}$ & $Y \sim x^{N/2}$ \\
&&&\\ \hline &&&\\
$\nu(\Sigma): \#$ of branch points&&& $N$ punctures at pairwise \\
i.e. where $Y=0$&$4N$ & $2N$ & contracted $2N$ branch points\\
&&&\\ \hline &&&\\
genus $g(\Sigma) = \frac{\nu(\Sigma)-2}{2}+2$ &
$2N+1$&$N+1$& $g=1$ torus with $N$ punctures\\
&&&\\ \hline
\end{tabular}


\hspace{-1cm}
\begin{tabular}{|c|c|c|c|}
\hline &&&\\
&$\frac{(1,\ldots,x^{N})dx}{Yy}$
&$\frac{(1,\ldots,x^{[N/2]})dx}{\left(w-\frac{1}{w}\right)y}$
&$\frac{(1,\ldots,x^{[N/2]})dx}{wy}$\\
&&&\\
holomorphic differentials on $\Sigma$&
$\frac{(1,\ldots,x^{N-2})dx}{Y}$
&$\frac{(1,\ldots,x^{[(N-3)/2]})dx}{w-\frac{1}{w}}$
&  $\frac{(1,\ldots,x^{[(N-3)/2]})dx}{w}$\\
(or with simple poles in rational case)&&&\\
&$\frac{dx}{y}=-2dz$ &$\frac{dx}{y}=-2dz$ &  $\frac{dx}{y}=-2dz$ \\
&&&\\ \hline &&&\\
total number of differentials & $(N+1)+(N-1)+1$&\
$\left(\left[\frac{N}{2}\right]+1\right) +$&
$\ \left(\left[\frac{N}{2}\right]+1\right) +$\\
= $g(\Sigma) + \# {\rm marked\ points}$& $= 2N+1$ &
$+\left(\left[\frac{N-3}{2}\right]+1\right)+1=$ &
$+\left(\left[\frac{N-3}{2}\right]+1\right)+1=$ \\
&&\ $= N+1$&\ $= N+1$\\
&&&\\ \hline
&&&\\
symmetry of $\Sigma$ & $\mathbb{Z}_2:  \tilde z \leftrightarrow
-\tilde z,\ Y \leftrightarrow -Y$ & $\mathbb{Z}_2:  w
\leftrightarrow \frac{1}{w}
  $ & none \\
&&&\\ \hline &&&\\
&&&\\
holomorphic differentials, &$\frac{(1,\ldots,x^{N})dx}{Yy}$&
$\frac{(1,\ldots,x^{[N/2]})dx}{\left(w-\frac{1}{w}\right)y}$ &$\frac{(1,\ldots,x^{[N/2]})dx}{wy}$\\
{\em odd} under the symmetry &&&
\\
&$\frac{(1,\ldots,x^{N-2})dx}{Y}$
&$\frac{(1,\ldots,x^{[(N-3)/2]})dx}{w-\frac{1}{w}}$
&$\frac{(1,\ldots,x^{[(N-3)/2]})dx}{w}$ \\
&&&$\frac{dx}{y}=dz$\\
&&&\\ \hline &&&\\
$n_h''$: $\#$ of such differentials & $2N$ &$N$ &$N+1$\\
&&&\\ \hline &&&\\
$n_h'$: $\#$ of hol. differentials, &$n_h'=n_h''-1=2N-1$&
 & $n_h' = n_h'' - 1 = N$ \\
contributing to $\delta(dS)$ & one linear combination
&  $n_h' = n_h'' = N$ &$dz$ does not \\
&does not contribute && contribute\\
&&&\\ \hline &&&\\
$n_h = n'_h+1$ due to addition &&&\\
of $d\Omega_\tau = \frac{\partial (dS)}{\partial\tau}$
with jump & $2N$ & $N+1$ & $N+1$\\
&&&\\ \hline &&&\\
jump $\Delta_B (d\Omega_\tau)$ & $d\tilde z$&$\frac{dw}{w}$&$\frac{dw}{w}$\\
&&&\\ \hline
\end{tabular}

\hspace{-1.5cm}
\begin{tabular}{|c|c|c|c|} \hline
&&&\\
 & & $2$ simple poles & $N+1$ simple poles:\\
$d\tilde z$ & holomorphic &at $x=\infty$ ($z=0$)& at $x=\infty$ ($z=0$)\\
&&related by $w \leftrightarrow \frac{1}{w}$&and $N$ zeros of $P+yQ$\\
&&& or at $z=a_j$, $j=1,\dots,N$\\
&&&\\
\hline &&&\\
zeros of $d\tilde z$ (or of ${\cal H}'_z$) & $2N$ zeros of
$d\left( \frac{P+yQ}{R+yS}\right)$&
 $N+1$ zeros  of $d(P+yQ)$&
$N+1$ zeros \\
&doubled by $\tilde z \leftrightarrow -\tilde z$&
doubled by $w \leftrightarrow \frac{1}{w}$&of $d(P+yQ)$\\
&&&\\ \hline &&&\\
$n'_z$: number zeros of $d\tilde z$ & $4N
=2\left(g(\Sigma)-1\right)$ & $2N+2=2\left(g(\Sigma)-1\right) +
\#{\rm poles}$&
$N+1=\#{\rm poles}$\\
&&&\\ \hline &&&\\
$n_z$: $\#$ of critical points &&&\\
(zeros of $d\tilde z$ over symmetry) & $\frac{1}{2}n'_z = 2N$
&$\frac{1}{2}n_z'=N+1$&
$n_z'=N+1$\\
&&&\\
&&&\\ \hline\hline &&&\\
matching $n_m = n_h = n_z$ and& + & + & + \\
validity of WDVV equations&&&\\
&&&\\ \hline
\end{tabular}
\hspace{1.0cm}
\end{small}

\section{Discussion}

In this paper we have presented a generic check of the WDVV
equations arising as 6d SW prepotentials utilising the residue
formula. Further we have introduced a bi-elliptic generalization
for which the WDVV equations also hold. In particular this is one
of the very few cases known cases for which the curve is not
hyperelliptic. Our argument has however intensively used the
symmetry of the curves of the bi-elliptic family, reminiscent of
the hyperelliptic case.

In the most degenerate situation of the {\em perturbative} 6d
prepotentials the validity of the WDVV equations was recently
established in \cite{str}. The approach of the paper \cite{str}
was to extend the Landau-Ginzburg (LG) construction of the
superpotential to the torus \cite{Dub}. Now the complex torus is
endowed with the generating differential $dS^{\rm LG} = zdw$,
where $w$ from (\ref{6dc}) can be considered as the corresponding
superpotential. Equivalently the meromorphic function $w$ on the
complex torus gives the SW curve of a 6d theory. This situation
essentially parallels that of {\em perturbative} 4d SW theory and
the common LG model with a polynomial superpotential $W(X)$ on the
sphere with single marked point (or just on complex plane)
\cite{MMM,forms,MaWDVV}. In the common LG model the WDVV equations
follow from the polynomial ring modulo $W'(X)$, while in the
perturbative SW theory the corresponding algebra is isomorphic to
the ring of functions at the zeros of $W'(X)$. The nontrivial fact
for both cases is the existence of residue formulae for these {\em
different} functions - the LG and (perturbative) SW prepotentials
- which relate them to the structure constants of the isomorphic
algebras. Moreover it is known that the isomorphism with the LG
{\em algebra} holds in the SW theory {\em beyond} the perturbative
case, see \cite{MMM,MaWDVV}. Indeed one may write down explicit
formula expressing the isomorphic structure constants through the
third derivatives of the LG and {\em non-perturbative} SW
prepotentials, so relating one to the other
\cite{jap,luuk,MaWDVV}. In the light of these results the
extensions of the perturbative results of \cite{str} to the
non-perturbative regime are rather natural and perhaps not very
surprising.

We wish however to stress that the validity of the WDVV equations
established here does not depend on any of the extra requirements
intensively used in \cite{Dub}; the latter yielding
a well-known class of solutions to the WDVV equations related with
the simplest topological string theories. In particular, we have
already observed above that appearance of ``unity" in the basis of
functions and corresponding ``constant-metric" term in the
prepotential is accidental: both these features do not survive in
the non-perturbative and bi-elliptic cases, but the simple
counting argument (\ref{matching}) based on the residue formula
(\ref{residue}) still holds, as demonstrated in our tables.

We also note that there is some similarity between the families of
curves we have considered in this paper with another distinguished
family of SW curves: the softly-broken \4N theory described by the
Calogero-Moser integrable systems \cite{CM} in spirit of the
correspondence of \cite{GKMMM}. The SW curves for this family also
cover a complex torus and, as in the cases examined here, the
prepotential does not satisfy the WDVV equations as a function of
the SW periods alone \cite{more,MaWDVV}. However, in the
Calogero-Moser case the counting argument shows \cite{MaWDVV},
that one must add at least $(N-2)$ parameters and differentials,
unlike the single differential with jump and corresponding modulus
of the torus needed for the whole bi-elliptic family considered in
this paper.

Finally, it would be interesting to extend analysis of this paper
to the curves of the {\em double}-elliptic family, the simplest
examples of which were considered in \cite{dell,dellMM,bhgt}. Compared to the
bi-elliptic curves in the simplest double-elliptic case we have
\begin{equation}
\label{dell}
\Sigma_{\rm dell}:\ \ \ {\cal H}(z,\tilde z) =
\alpha(z){\rm cn} \left(\beta(z)\tilde
z\left|\frac{\alpha(z)}{\beta(z)}\tilde k\right.\right) - E =0,
\end{equation}
with $\alpha^2 = 1-c\wp(z) =1 -cx$ and $\beta^2 = 1 - {\tilde
k}^2c\wp(z) = 1 - {\tilde k}^2cx$. Here there is a non-trivial
periodicity in ${\tilde z}$-variable, with the periods themselves
depending on $z$. For the double-elliptic curve \rf{dell} one gets
\begin{equation}
\label{vardell}
\left.\delta(dS)\right|_{\delta\tilde z=0} =
\delta{\cal H}\frac{d\tilde z}{{\cal H}'_z} \propto
\frac{\delta{\cal H}dz}{\sqrt{x - U}} \propto \frac{\delta{\cal
H}dx}{\sqrt{(x - U)\prod_{i=1}^3(x-e_i)}}
\end{equation}
with $U = \frac{1-E^2}{c}$. Expression \rf{vardell} may be thought
of either as a differential on a double cover of the
$(z,\tau)$-torus, ramified at the two solutions of $x-U=0$, or as
a differential on a curve of genus $g\left(\Sigma_{\rm
dell}\right) = 2$. The naive number of holomorphic differentials
is then $n_z = 2\left(g\left(\Sigma_{\rm dell}\right) -1\right) =
2$, and this is not enough to analyse the validity of the WDVV
equations, even upon restricting number of moduli to $n_h=2$. We
remark, however, that the derivation of \rf{vardell} contains many
surprising cancellations: for example, due to $\beta^2-{\tilde
k}^2\alpha^2= {\rm const}$, it means that one can look for the
nontrivial double-elliptic solutions to the WDVV equations for the
families of curves similar to \rf{dell}. We will return to this
problem elsewhere.

\section*{Acknowledgements}

We are grateful to I.~Krichever, A.~Rosly and I.~A.~B. Strachan
for discussion. The work was partially supported by RFBR grants
03-02-17373 (A.Mar.), 04-01-16646 (A.Mir.), 04-02-16880 (A.Mor.),
the grants for support of Scientific Schools LSS-4401.2006.2
(A.Mar.) and LSS-8004.2006.2 (A.Mir., A.Mor.), the NWO projects
047.017.2004.015 (A.Mar.) and 047.011.2004.026 (A.Mir., A.Mor.),
the INTAS grant 05-1000008-7865 and the ANR-05-BLAN-0029-01
project ``Geometry and Integrability in Mathematical Physics"
(A.Mar, A.Mir, A.Mor). H.W.B. and A.Mar. would like to thank the
Max Planck Institute for Mathematics where this work was
completed, and the Edinburgh Mathematical Society for their
support.


\begin{thebibliography}{7799}

\bibitem{WDVV}
E.~Witten, Nucl. Phys. {\bf B340} (1990) 281;\\
R.~Dijkgraaf, H.~Verlinde and E.~Verlinde, Nucl. Phys. {\bf B352}
(1991) 59.
%
\bibitem{MMM}
A.~Marshakov, A.~Mironov and A.~Morozov, Phys. Lett. {\bf B389}
(1996) 43,  [arXiv:hep-th/9607109].
%
\bibitem{Dub}
B.~Dubrovin, 
in \emph{Integrable systems and quantum groups} (Montecatini Terme,
1993), 120--348, Lecture Notes in Math., 1620, Springer, Berlin,
1996,
  [arXiv:hep-th/9407018].
%
\bibitem{forms}
  A.~Marshakov, A.~Mironov and A.~Morozov,
  Mod.\ Phys.\ Lett.\ A {\bf 12} (1997) 773
  [arXiv:hep-th/9701014].
%
\bibitem{more}
  A.~Marshakov, A.~Mironov and A.~Morozov,
  Int.\ J.\ Mod.\ Phys.\ A {\bf 15} (2000) 1157
  [arXiv:hep-th/9701123].
%
\bibitem{MM} A.~Morozov, Phys.Lett. {\bf B427} (1998) 93-96
[arXiv:hep-th/9711194];\\
A.~Mironov and A.~Morozov, Phys.Lett. {\bf B424} (1998) 48-52
[arXiv:hep-th/9712177].
%
\bibitem{dWMa}
B.~de Wit and A.~Marshakov,
  Theor.\ Math.\ Phys.\  {\bf 129} (2001) 1504,
  [arXiv:hep-th/0105289].
%
\bibitem{KriW}I.~Krichever,
Commun. Pure. Appl. Math. {\bf 47} (1992) 437, [arXiv:hep-th/9205110].
%
\bibitem{mir} A.~Mironov, the review in the second book of [\ref{bk}],
[arXiv:hep-th/9903088].
%
\bibitem{BMRWZ} A.Boyarsky, A.Marshakov, O.Ruchayskiy, P.Wiegmann and
A.Zabrodin,
Phys.Lett. {\bf 515B} (2001) 483-492, [arXiv:hep-th/0105260].
%
\bibitem{MaWDVV}
A. Marshakov, in {\em Supersymmetry and Unification of Fundamental Interactions}, D.~Kazakov and A.~Gladyshev (Eds.),
(World Scientific, Singapore, 2002), pp 370-381, [arXiv:hep-th/0108023];\\
Theor.Math.Phys. {\bf 132} (2002) 895, [arXiv:hep-th/0201267].
%
\bibitem{SW}
N.Seiberg and E.Witten, Nucl. Phys. {\bf B426} (1994) 19;
[arXiv:hep-th/9407087].
%
\bibitem{SWbooks}\label{bk}
A.~Marshakov, {\em Seiberg-Witten Theory and Integrable Systems},
(World Scientific, Singapore, 1999);
\\
{\em Integrability: The Seiberg-Witten and Whitham Equations},
H.W.~Braden and I.~Krichever (Eds.), (Gordon and Breach, 2000).
%
\bibitem{GMrev}
see, for example:\\
A.~Marshakov,
  Theor.\ Math.\ Phys.\  {\bf 112} (1997) 791
  [arXiv:hep-th/9702083];\\
  R.~Donagi, in Surv. Differ. Geom., IV, Int. Press, Boston, MA, 1998. arXiv:alg-geom/9705010;\\
  I.~M.~Krichever and D.~H.~Phong, in Surv. Differ. Geom., IV, Int. Press, Boston, MA, 1998.
  arXiv:hep-th/9708170;\\
A.~Gorsky and A.~Mironov,
arXiv:hep-th/0011197.
%
\bibitem{XYZ}
A.~Gorsky, A.~Marshakov, A.~Mironov and A.~Morozov,
  arXiv:hep-th/9604078.
%
\bibitem{MaMi}
A.~Marshakov and A.~Mironov,
  Nucl.\ Phys.\ B {\bf 518} (1998) 59
  [arXiv:hep-th/9711156].
%
\bibitem{str}
A.~Riley and I.~Strachan, arXiv:math-ph/0511048.
%
\bibitem{dell}
H.W.~Braden, A.~Marshakov, A.~Mironov and A.~Morozov,
  Nucl. Phys. {\bf B573} (2000) 553-572, [arXiv:hep-th/9906240].\
%
\bibitem{Ino}
V.~Inozemtsev, Comm. Math. Phys. {\bf 121} (1989) 629.
%
\bibitem{IM2}
H.~Itoyama and A.~Morozov,
  Nucl.\ Phys.\ B {\bf 491} (1997) 529
  [arXiv:hep-th/9512161].
%
\bibitem{jap}
K.~Ito and S.-K.~Yang, Phys. Lett. {\bf B433} (1998) 56-62,
[arXiv:hep-th/9803126].
%
\bibitem{luuk}
L.~Hoevenaars and R.~Martini, Lett.~Math.~Phys. {\bf 57} (2001)
175-183, [arXiv:hep-th/0102190].
%
\bibitem{CM}
R.Donagi and E.Witten, Nucl.Phys. {\bf B460} (1996) 299; [arXiv:hep-th/9510101];\\
E.Martinec,
Phys.Lett. {\bf B367} (1996) 91-96; [arXiv:hep-th/9510204];\\
A.Gorsky and A.Marshakov,  Phys.Lett. {\bf B375} (1996) 127, [arXiv:hep-th/9510224];\\
H.Itoyama and A.Morozov, Nucl.Phys. {\bf B477} (1996) 855;
[arXiv:hep-th/9511126]; [arXiv:hep-th/9601168];\\
 E.~D'Hoker and D.~H.~Phong,
  Nucl.\ Phys.\ B {\bf 513} (1998) 405
  [arXiv:hep-th/9709053].
%
\bibitem{GKMMM}
A.Gorsky, I.Krichever, A.Marshakov, A.Mironov and A.Morozov, Phys.
Lett. {\bf B355} (1995) 466; [arXiv:hep-th/9505035].
%
\bibitem{dellMM}
A.~Mironov and A.~Morozov, Phys.Lett. {\bf B475} (2000) 71-76,
[arXiv:hep-th/9912088];\\
A.~Mironov and A.~Morozov, arXiv:hep-th/0001168.
%
\bibitem{bhgt}
H. W. Braden and T. Hollowood,
JHEP {\bf12} 023, 2003 [arXiv:hep-th/0311024].
%

\end{thebibliography}
\end{document}